\documentclass[12pt, prd, showpacs]{revtex4}
\usepackage{amssymb}
\usepackage{amsmath}

\input{tcilatex}

\begin{document}

\title{Acceleration of particles by black holes as a result of deceleration:
ultimate manifestation of kinematic nature of BSW effect}
\author{Oleg B. Zaslavskii}
\affiliation{Kharkov V.N. Karazin National University, 4 Svoboda Square, Kharkov, 61077,
Ukraine}
\email{zaslav@ukr.net}

\begin{abstract}
The recently discovered so-called BSW effect consists in the unbound growth
of the energy $E_{c.m.}$ in the centre of mass frame of two colliding
particles near the black hole horizon. We consider a new type of the
corresponding scenario when one of two particles ("critical") remains at
rest near the horizon of the charged near-extremal black hole due to balance
between the attractive and repulsion forces. The other one hits it with a
speed close to that of light. This scenario shows in a most pronounced way
the kinematic nature of the BSW effect. In the extremal limit, one would
gain formally infinite $E_{c.m.}$ but this does not happen since it would
have require the critical massive particle to remain at rest on the null
horizon surface that is impossible. We also discuss the BSW effect in the
metric of the extremal Reissner-Nordstr\"{o}m black hole when the critical
particle remains at rest near the horizon.
\end{abstract}

\maketitle



\section{Introduction}

The recent finding of the effect of the unbound growth of the energy $%
E_{c.m.}$in the centre of mass frame due to collisions of particles near the
black \ hole horizon \cite{ban} (BSW effect) attracts now much attention.
Both manifestations of this effect in different situation are being studied
in detail and also the very nature of the effect itself is under
investigation. It was observed in \cite{k} that the underlying physical
reason of the BSW effect can be explained in kinematic terms. Namely, it
turns out that, roughly speaking, a rapid particle collides with a slow one
near the horizon, this leads to the growth of the relative velocity and, as
a result, to the unbound growth of the corresponding Lorentz gamma - factor,
so the energy $E_{c.m.}$ becomes unbound near the horizon. This general
circumstance was also confirmed in thorough analysis of the BSW effect in
the Kerr metric \cite{gpp}. Nonetheless, some doubts remain concerning the
possibility to give an alternative explanation. If something is being
accelerated to unbound energies, one is tempted to ask, what source does
this, and what is the "physical" underlying reason of such an effect.

The aim of the present work is to reveal the kinematic nature of the BSW
effect in the most pronounced way. To this end, we consider the situation
when one of two colliding particles is motionless while the other one moves
(as usual) with a finite energy in the frame of a distant observer. In a
sense, this is the ultimate and clear manifestation of the kinematic nature
of the effect under discussion that does not require to search for further
hidden dynamic factors. The model which we discuss shows the key issue as
clear as possible: the role of gravitation in producing the BSW effect of
the unbound growth of $E_{c.m.}$ ("acceleration of particles") consists not
in acceleration but in \textit{deceleration }of one of two particles (in the
sense that its velocity is reduced to zero)!

To achieve our goal, we consider the spherically symmetric metric of a
charged black hole that admits the equilibrium of a particle that remains
motionless. In other words, we want to balance the gravitation force by
electrical repulsion. Apart from this, it is important that such a point be
located in the vicinity of the horizon. For definiteness, we consider the
innermost stable equilibrium point which is the counterpart of the innermost
stable circular orbit for the Kerr metric \cite{72}. Such orbits were
discussed recently due to their potential astrophysical significance \cite%
{kerr}, \cite{spiral}. (See their generalization to "dirty" rotating black
holes \cite{d}.) There exists also their analog in the magnetic field where
the BSW effect was studied recently in \cite{fr}.

The simplest choice would seem to be the Reissner-Nordtr\"{o}m (RN) black
hole but for this metric the "orbit" with the required properties exists for
indifferent equilibrium only (see Sec. V below). Therefore, for the analog
of inner stable orbits we take the charged black hole with nonzero
cosmological constant $\Lambda $. It turns out that it is required that $%
\Lambda <0$, so we mainly deal with the Reissner-Nordtr\"{o}m - anti-de
Sitter one (RN-AdS) which is sufficient for our purposes. It is also worth
noting that interest to black holes with the cosmological constant $\Lambda
<0$ revived in recent years due to AdS/CFT correspondence \cite{mald}. In
addition, we consider also another type of \textquotedblleft
orbit\textquotedblright\ -- a particle in the state of indifferent
equilibrium in the metric of the extremal Reissner--Nordstr\"{o}m black hole.

\section{Equations of motion}

Consider the space-time describing a charged black hole with the
cosmological constant.\ Its metric can be written as 
\begin{equation}
ds^{2}=-fdt^{2}+\frac{dr^{2}}{f}+r^{2}d\Omega ^{2},  \label{m}
\end{equation}%
\begin{equation}
f=1-\frac{2m}{r}+\frac{Q^{2}}{r^{2}}-\frac{\Lambda r^{2}}{3}\text{.}
\label{fr}
\end{equation}%
Throughout the Letter we assume that the fundamental constants $G=c=$%
h{\hskip-.2em}\llap{\protect\rule[1.1ex]{.325em}{.1ex}}{\hskip.2em}
$=1$. The horizon lies at $r=r_{+}$ where $f(r_{+})=0$. The electric
potential 
\begin{equation}
\varphi =\frac{Q}{r}+C  \label{phi}
\end{equation}%
where the $C$ is the constant of integration. It is assumed that we work in
the gauge where the only nonvanishing component $A_{0}=-\varphi $. For the
asymptotically flat case, say, for the Reissner-Nordstr\"{o}m or Kerr-Newman
metric, it is usually chosen $C=0$ to have $\varphi =0$ at infinity (see,
e.g., eq. 3.63 of \cite{fn}). In the absence of asymptotic flatness, its
choice becomes conditional. It is worth stressing that $\ $physically
relevant quantities contain not the potential itself but the difference with
respect to some reference point (infinity or horizon). For example, in black
hole thermodynamics, the potential enters the action in the form $\frac{%
\varphi (r)-\varphi (r_{+})}{\sqrt{f}}$ that is nonsingular at the horizon
(see eq. 4.15 of \cite{rn}). In equations of motion (see below) only the
combination $E-q\varphi $ appears where $q$ is the particle's charge. If we
change the potential according to $\varphi \rightarrow \varphi +C$, the
corresponding shift in the energy $E\rightarrow E+qC$. For convenience, we
choose $C=0$ in (\ref{phi}).

We restrict ourselves by radial motion since this case is the most
interesting in the context under discussion. As is known, under the presence
of the electromagnetic field, dynamics of the system is described by the
generalized momentum $P_{\mu }$ related to the kinematic one $p_{\mu
}=mu^{\mu }$ by the relation $p_{\mu }=P_{\mu }-qA_{\mu }$ where $u^{\mu }=%
\frac{dx^{\mu }}{d\tau }$ is the four-velocity of a test massive particle, $%
\tau $ is the proper time, $A_{\mu }$ is the vector potential. Due to
staticity, the energy $E=-P_{0}$ of a particle moving in this metric is
conserved, $P_{0}$ is the time component of the generalized momentum $P_{\mu
}$. Then, using also the relation $u^{0}=g^{00}u_{0}$, we obtain (dot
denotes the derivative with respect to the proper time $\tau $)%
\begin{equation}
\dot{t}=u^{0}=\frac{X}{mf},  \label{t}
\end{equation}%
\begin{equation}
X=E-q\varphi \text{.}  \label{x}
\end{equation}%
We assume that $\dot{t}>0$, so that $E-q\varphi >0$. 
\begin{equation}
m^{2}\dot{r}^{2}=-V_{eff}=X^{2}-m^{2}f\text{. }  \label{rv}
\end{equation}%
Now, we are interested in equilibrium solutions $r=r_{0}=const$, 
\begin{equation}
V_{eff}(r_{0})=0.  \label{v}
\end{equation}%
Additionally, we require that they possess the following properties: (i) $%
r_{0}$ is a perpetual turning point, (ii) it lies near the horizon, $%
r_{+}\rightarrow r_{0}$. Condition (i) means that, in addition to (\ref{v}),
equation%
\begin{equation}
V_{eff}^{\prime }(r_{0})=0  \label{v0}
\end{equation}%
should hold. Eqs. (\ref{v}), (\ref{v0}) ensure that not only $\dot{r}$ but
also all higher derivatives vanish.

It follows from (\ref{rv}), (\ref{v}) that for a particle with $\dot{r}=0$, 
\begin{equation}
X(r_{0})=m\sqrt{f(r_{0})}\text{.}  \label{zxi}
\end{equation}%
It is instructive to elucidate for which types of black holes equations (\ref%
{v}) and (\ref{v0}) are self-consistent near the horizon, so that
equilibrium points exist there in agreement with requirement (ii). Physical
motivation for considering this requirement comes from our main goal -
investigation of the BSW effect since this effect occurs just in the
vicinity of the horizon.

If we \ take the derivative of the effective potential $V_{eff}$ in eq. (\ref%
{rv}) and take into account also the relation (\ref{v}), we obtain%
\begin{equation}
-\frac{1}{2}V_{eff}^{\prime }(r_{0})=m\sqrt{f(r_{0})}\frac{qQ}{r_{0}^{2}}-%
\frac{m^{2}}{2}f^{\prime }(r_{0}).  \label{v'}
\end{equation}

Let us consider the limit $r_{0}\rightarrow r_{+}$, so $f(r_{0})\rightarrow
0 $. Then, it follows from (\ref{v'}) that $V_{eff}^{\prime
}(r_{0})\rightarrow -m^{2}\kappa $ where we used the fact for the metric (%
\ref{m}) $\kappa =\frac{1}{2}f^{\prime }(r_{+})$. Thus if $\kappa \neq 0$,
eq. (\ref{v0}) cannot be satisfied in the horizon limit. Therefore, for
nonextremal black holes the equilibrium points cannot exist near the horizon
(although they can exist elsewhere at a finite distance from the horizon).
This generalizes previous observations \cite{gpp}, \cite{d} made for
rotating black holes. However, if $\kappa \rightarrow 0$, the equilibrium
points close to the horizon do exist as will be shown below.

\section{Properties of equilibrium point}

For the Kerr metric \cite{72} and, in general, for axially-symmetric
rotating black holes \cite{d}, there are so-called innermost stable orbits
(ISCO) which correspond to the threshold of stability. We consider now their
analogs in our case, so we must add to (\ref{v}) and (\ref{v'}), also
equation%
\begin{equation}
V_{eff}^{^{\prime \prime }}(r_{0})=0.  \label{v''}
\end{equation}%
For brevity, we will call this an innermost stable equilibrium point (ISEP).

We are interested in the near-horizon region where we can expand $f$ in the
Taylor series with respect to $x=r_{0}-r_{+}$:%
\begin{equation}
f=2\kappa x+Dx^{2}+Cx^{3}...\text{.}  \label{f}
\end{equation}

From now on, we assume that $\kappa $ is a small parameter, so a black hole
is a near-extremal. Then, this leads to an interplay between two small
quantities $\kappa $ and $x$. We assume the condition%
\begin{equation}
\kappa \ll Dx
\end{equation}%
which one can check a posteriori that for the solutions obtained.

Then, the procedure for the description of the equilibrium points is
mathematically similar to that for the description of circular orbits in the
background of rotating black holes \cite{d}. In both cases, we are
interested in solutions for which $\dot{r}=0$ and which are on the threshold
of stability. Therefore, I omit technical details (which are connected with
simple but rather cumbersome calculations) and give the main results of eqs.
(\ref{v}), (\ref{v0}), (\ref{v''}).

It turns out that%
\begin{equation}
x^{3}\approx H^{3}\kappa ^{2}\text{,}  \label{xh}
\end{equation}%
where%
\begin{equation}
H^{3}=\frac{3r_{+}^{3}}{4(-\Lambda )(1-2\Lambda r_{+}^{2})}  \label{h}
\end{equation}%
and the constants in (\ref{f}) 
\begin{equation}
D=\frac{1}{r_{+}^{2}}-2\Lambda \text{,}  \label{D}
\end{equation}%
\begin{equation}
C=-\frac{2}{r_{+}^{3}}+\frac{8}{3}\frac{\Lambda }{r_{+}}.
\end{equation}

As in the extremal limit $\kappa \rightarrow 0$ we must have $f>0$ in the
vicninity of the horizon from the outside, the coefficient $D>0$. Then, in
combination with $H>0$, this entails that $\Lambda <0$.

By substitution of (\ref{xh}) into (\ref{f}) we obtain%
\begin{equation}
\sqrt{f}\approx \left( \frac{3}{4}\right) ^{1/3}(-\Lambda
)^{-1/3}(1-2\Lambda r_{+}^{2})^{1/6}\kappa ^{2/3}\text{.}  \label{fk}
\end{equation}

Although $r_{0}\rightarrow r_{+}$, the proper distance $l=\int \frac{dr}{%
\sqrt{f}}$ between the particle and the horizon does not vanish and,
moreover, it grows unbound when $\kappa \rightarrow 0$:%
\begin{equation}
l\approx \frac{1}{\sqrt{D}}\ln \frac{x_{0}}{\kappa }\approx \frac{1}{3}\ln 
\frac{1}{\kappa }\text{.}
\end{equation}

This is in full analogy with the rotating case \cite{72}, \cite{d}.

\section{Collisions with unbound energy}

Now, we consider the collision of two particles. To avoid unnecessary
complications due to possible Coulomb repulsion between particles having the
charge of the same sign, we can assume that the particle falling towards a
black hole is neutral. Assuming, for simplicity, that both particles have
the same mass $m$, the energy is given by the formula \cite{jl}%
\begin{equation}
\frac{E_{c.m.}^{2}}{2m^{2}}=1+\frac{X_{1}X_{2}-Z_{1}Z_{2}}{fm^{2}}
\label{eq}
\end{equation}%
where 
\begin{equation}
Z_{i}=\sqrt{X_{i}^{2}-m^{2}f}\text{, }i=1,2\text{.}
\end{equation}%
Let, for definiteness, a motionless particle have $i=1$. The unbound growth
of the energy $E_{c.m.}$ occurs if one particle has on the horizon $\left(
X_{1}\right) _{+}=0$ (we call it critical) whereas for the other one $\left(
X_{2}\right) _{+}\neq 0$ (we call it usual) - see \cite{jl} for details.

In our case, $Z_{1}=0$ according to eq. (\ref{zxi}), so the formula
simplifies:%
\begin{equation}
\frac{E_{c.m.}^{2}}{2m^{2}}=1+\frac{X_{2}}{m\sqrt{f}}.  \label{ef}
\end{equation}

Particle 1 has $X_{1}\sim \sqrt{f}$, so it is near-critical near the horizon 
$f\rightarrow 0$ which is just the case we are dealing with. Then, we obtain
for the collision near the horizon that the energy has the form%
\begin{equation}
E_{c.m.}\approx \sqrt{2mX_{2}}A\kappa ^{-\frac{1}{3}}  \label{ek}
\end{equation}%
where it follows from (\ref{fk}) that 
\begin{equation}
\text{ }A=\left( \frac{4}{3}\right) ^{1/6}(-\Lambda )^{1/6}(1-2\Lambda
r_{+}^{2})^{-1/12}\text{.}
\end{equation}

The dependence $E_{c.m.}\sim \kappa ^{-1/3}$ is similar to that for rotating
black holes (cf. eq. 5.1 of \cite{kerr} and eq. 89 of \cite{d}).

\section{Degenerate case: BSW effect for a particle at rest in the extremal
Reissner-Nordstr\"{o}m metric}

In investigation of the BSW effect, one is led to deal with different
limiting transitions: $r_{0}\rightarrow r_{+}$, $\left( X_{1}\right)
_{+}\rightarrow 0$ that requires certain care. It was demonstrated earlier
(see eqs. 11 and 15 of \cite{prd} and eqs. 8, 10 of \cite{jl}) that these
limits do commute and give $E_{c.m.}\rightarrow \infty $ in both cases for
collisions of particles moving towards the horizon of an extremal black
hole. In our case there are two distinctions from the aforementioned
situation: (i) a black hole is nonextremal with small but nonzero $\kappa $,
(ii) the point of collision cannot be considered as an independent parameter
since it coincides with the equilibrium point of particle 1 whose location $%
r_{0}$ is controlled by $\kappa $ according to (\ref{xh}), (\ref{h}).
Therefore, one cannot make permutation between $r_{0}\rightarrow r_{+}$ and $%
\kappa \rightarrow 0$ that represents now a self-consistent indivisible
procedure. Meanwhile, one may ask what happens to the points of equilibrium
if the limit $\kappa =0$ is taken from the very beginning that corresponds
to an extremal black hole.

Formally, $r_{0}\rightarrow r_{+}$ in this limit. However, on the horizon
which is light-like surface, the time-like trajectories cannot exist, so the
solution $r_{0}=r_{+}$ for it is fake (see the detailed analysis in Sec. III
C of \cite{kerr}). The real trajectory is not strictly static and
asymptotically approaches the horizon \cite{ted}, \cite{gp}, \cite{prd}, 
\cite{kerr}. It would seem that in such circumstances the questions about
ISEP in the near-horizon region do not make sense at all. Nonetheless, there
is an exceptional case when ISEP degenerates into points of indifferent
equilibrium. This happens just for extremal Reinssner-Nordstrom black holes.
Then, in eq. (\ref{fr}), $Q=M=r_{+}$, $\Lambda =0$. By direct check, it is
easy to see that eqs. (\ref{v}), (\ref{v0}) lead to the consequences that $%
E=m=q$. But for these values of particle's parameters, the effective
potential $V_{eff}=0$ for any $r$. Actually, this means that a particle can
be at rest at any position $r_{0}$ due to balance between gravitational
attraction and electric repulsion, so equilibrium is indifferent. (More on
the properties of equilibrium in the Reissner-Nordstr\"{o}m space-time can
be found in \cite{bon}).

Now, eq. (\ref{ef})\ gives us an exact expression%
\begin{equation}
\frac{E_{c.m.}^{2}}{2m^{2}}=1+\frac{X_{2}}{1-\frac{r_{0}}{r_{+}}}
\end{equation}%
for an arbitrary $r_{0}$. Thus we can see that for $r_{0}\rightarrow r_{+}$, 
$E_{c.m.}\sim (1-\frac{r_{0}}{r_{+}})^{-1/2}$. In other words, we place
particle 1 in any point at rest and inject another particle from the outside
(say, from infinity). When the location of particle 1 approaches the
horizon, $E_{c.m.}\rightarrow \infty $, so we again obtain the BSW effect.

\section{Kinematic censorship}

There is one more question concerning the possibility to take the limit $%
\kappa \rightarrow 0$. It follows from our formulas for the BSW effect at
ISEPs that for any small but nonzero $\kappa $ the energy $E_{c.m.}$ is
large but finite. Can one simply take the value $\kappa =0$ from the very
beginning and gain an infinite energy? In any real physical event the actual
energy that can be released must be finite. With respect to collisions of
particles, it can be named "kinematic censorship". Therefore, the energy $%
E_{c.m.}$ can be as large as one likes but it cannot be literally infinite.
To understand, how this kinematic censorship is realized in our case, one
should take into account explanations from the previous Section. We would
like to stress it once again that the "orbit" $r_{0}=r_{+}$ to which
formally tends the ISEP is not suitable since a trajectory of a massive
particle cannot lie on the light-like horizon surface. In the example with
indifferent equilibrium in the extremal Reinssner-Nordstrom background, the
situation is even more clear: we can place particle 1 at any position $%
r_{0}>r_{+}$ which is as close to $r_{0}$ as one likes but it cannot
coincide with $r_{0}$ nonetheless.

\section{Role of gravity in BSW effect}

The results obtained concern charged black holes and represent a counterpart
of those for the circular orbits in the background of rotating black holes 
\cite{kerr}, \cite{d}. Meanwhile, the basic point does not have an analogue
in the case of the Kerr metric. Indeed, the circular orbits in the
near-horizon region of the Kerr black hole lie in the ergosphere, so
equilibrium is not possible there. The perpetual turning point in \cite{72}
is related to circular orbit, so a particle necessarily has the nonzero
angular velocity with respect to a distant observer. Meanwhile, in the
present work a particle located near the horizon has both zero velocity and
zero angular momentum.

The BSW effect with participation of such a particle sharpens and reveals
its kinematic nature. Naively, one could think that the particle is
accelerated during the infall into the black hole and the problem seemed to
be to explain why and how this happens. However, we see that the real
picture is quite different. One of two colliding particles is kept fixed to
remain at rest near the horizon while the other one starts its motion from
the outside. As the all process is essentially nonlocal, one cannot
attribute the growth of relative energy to the action of force that exerts
on some united object. Moreover, in the collision under discussion particle
2 that comes from the outside is typical ("usual") in that it has an
arbitrary finite individual energy and zero charge, so the relation $X_{+}=0$
typical of the critical particle is not satisfied for it. As a result, its
velocity in the static frame approaches that of light \cite{k}, and this
becomes true for any such a particle. As a result, the relative velocity of
two usual particles remains finite and the BSW effect is impossible (see 
\cite{k} for details in general and analysis for the\ Kerr metric in \cite%
{gpp}). To gain the BSW effect, one should select such a particle that
approaches the velocity less than that of light near the horizon. In the
present case, it is the particle with literally zero velocity that remains
zero all the time before collision. In other words, the role of gravitation
consists here not in acceleration of particle 1 but in keeping it in rest
due to balancing electric repulsion, so the eventual outcome of the unbound
energy $E_{c.m.}$ is obtained in a\ sense as a consequence of deceleration
or, more precisely, arrest of one of particles! Moreover, in the case of the
Reissner-Nordstr\"{o}m-anti-de Sitter black hole the combined action of
gravitation, electricity and the cosmological term is to arrest particle 1
in such a way that it remains there on the threshold of stability. (One can
take a particle with $V_{eff}^{\prime \prime }>0$ instead of (\ref{v''}) to
have it strictly stable with respect to radial displacement.) Another
example with the same qualitative features is indifferent equilibrium of a
charged particle in the metric of the extremal Reinssner-Nordstr\"{o}m black
hole. It is this particle that remains at rest is now a critical one. It is
in accord with the general principle that for the BSW effect to occur, one
of particles should be critical and the other one should be usual \cite{k}, 
\cite{prd}.

Thus the case considered above shows in a most pronounced way that the
kinematic nature of the BSW enables one to obtain unbound energies because
of different action of gravity on essentially different (in a kinematic
sense) kinds of particles.

It is worth noting that in \cite{gpp} the collision between the infalling
particle and the particle at rest was discussed (after eq. 34). Meanwhile,
there is a qualitative difference between both situations. In \cite{gpp},
the particle was kept at rest in the Schwazrschild background that was
possible "by hand" only and required an almost infinite force near the
horizon. Meanwhile, in the example considered in our work, one can check
that the value of acceleration $a$ if finite. Indeed, one can calculate the
scalar $a^{2}=a_{\mu }a^{\mu }$ where $a_{\mu }=u_{\mu ;\nu }u^{\mu }$,
semicolon denotes the covariant derivative. It is easy to find that for the
metric (\ref{m}), $a^{2}=\frac{\left( f^{\prime }\right) ^{2}}{4f}\approx 
\frac{1-2\Lambda r_{+}^{2}}{r_{+}^{2}}$ is finite due to the near-extremal
character of a black hole. Thus gravity in combination with electric
repulsion and the cosmological constant, ensures the BSW effect on a
motionless particle in a self-consistent way. In doing so, the particle
waiting at rest plays the role of a target which is hit by the infalling
particle.

\section{Conclusion and perspectives}

The type of the BSW effect discussed in the present Letter is somewhat
different from the original one considered in the pioneering paper \cite{ban}%
. It makes the kinematic nature of the BSW effect especially pronounced.
Inclusion of backreaction and radiation into general scheme can change the
details of the effect significantly \cite{bert}, \cite{ted} but one can
expect that the main qualitative features of the BSW effect still persist
just due to its kinematic nature. Moreover, the fact that in the scenario
discussed in the present work, a near-critical or critical particle remains
at rest, suggests that for it gravitational radiation (mentioned in
aforementioned papers as a factor acting against the BSW effect) is absent
now at all. The role of backreaction on the metric is more subtle but,
anyway, the scenario of collisions considered in the present paper
simplifies the picture and can be useful for further analysis. More detailed
study of the generic BSW effect with account for all these factors is a
nontrivial problem that needs separate treatment.

\end{document}